\documentstyle[epsf,eqsecnum,floats,preprint,aps]{revtex}

\tighten

\input epsf
\begin{document}

\newcommand{\be}{\begin{equation}}
\newcommand{\ee}{\end{equation}}
\newcommand{\bea}{\begin{eqnarray}}
\newcommand{\eea}{\end{eqnarray}}
\newcommand{\PSbox}[3]{\mbox{\rule{0in}{#3}\includegraphics{#1}\hspace{#2}}}
\newcommand{\modified}[1]{{\it #1}}

\def\5M{M^3_{(5)}}
\def\4M{M^2_{(4)}}
\def\mpsq{M_{\rm P}^2}
\def\om{\Omega_m}
\def\omt{\Omega_m^0}

\overfullrule=0pt
\def\Int{\int_{r_H}^\infty}
\def\d{\partial}
\def\e{\epsilon}
\def\M{{\cal M}}
\def\high{\vphantom{\Biggl(}\displaystyle}
\catcode`@=11
\def\@versim#1#2{\lower.7\p@\vbox{\baselineskip\z@skip\lineskip-.5\p@
    \ialign{$\m@th#1\hfil##\hfil$\crcr#2\crcr\sim\crcr}}}
\def\simge{\mathrel{\mathpalette\@versim>}} %
\def\simle{\mathrel{\mathpalette\@versim<}} %
\def\sun{\hbox{$\odot$}}
\catcode`@=12 

\rightline{CERN--PH--TH/2004--007}
\rightline{astro-ph/0401515}
\vskip 2cm

\setcounter{footnote}{0}

\begin{center}
\large{\bf Probing Newton's Constant on Vast Scales:\\
DGP Gravity, Cosmic Acceleration and Large Scale Structure}
\ \\
\ \\
\normalsize{Arthur Lue,$^1$\footnote{E-mail: lue@cern.ch}
  Rom\'an Scoccimarro,$^2$\footnote{E-mail: rs123@nyu.edu}
  and Glenn D. Starkman$^1$\footnote{E-mail: glenn.starkman@cern.ch}}
\ \\
\ \\
${}^{1}$\small{\em
Center for Education and Research in Cosmology and Astrophysics\\
Department of Physics, Case Western Reserve University,
Cleveland, OH 44106--7079\\
and CERN Theory Division, CH--1211 Geneva 23, Switzerland}
\ \\
\ \\
${}^{2}$\small{\em Center for Cosmology and Particle Physics\\
Department of Physics, New York University, New York, NY 10003}

\end{center}

\begin{abstract}
\noindent
The nature of the fuel that drives today's cosmic acceleration is an
open and tantalizing mystery.  The braneworld theory of Dvali,
Gabadadze, and Porrati (DGP) provides a context where late-time
acceleration is driven not by some energy-momentum component (dark
energy), but rather is the manifestation of the excruciatingly slow
leakage of gravity off our four-dimensional world into an extra
dimension.  At the same time, DGP gravity alters the gravitational
force law in a specific and dramatic way at cosmologically accessible
scales.  We derive the DGP gravitational force law in a cosmological
setting for spherical perturbations at subhorizon scales and compute
the growth of large-scale structure. We find that a residual repulsive
force at large distances gives rise to a suppression of the growth of
density and velocity perturbations.  Explaining the cosmic
acceleration in this framework leads to a present day fluctuation
power spectrum normalization $\sigma_8\leq0.8$ at about the two-sigma
level, in contrast with observations. We discuss further theoretical
work necessary to go beyond our approximations to confirm these
results.
\end{abstract}

\setcounter{page}{0}
\thispagestyle{empty}
\maketitle

\eject

\vfill

\baselineskip 18pt plus 2pt minus 2pt

\section{Introduction}

The discovery of a contemporary cosmic
acceleration~\cite{Perlmutter:1998np,Riess:1998cb} is one of the most
profound scientific observations of the 20th century.  We are now
challenged to answer the open and tantalizing question of what drives
that acceleration.  \modified{While a conventional explanation exists
(i.e. dark energy -- some new negative-pressure energy-momentum
component)},
an intriguing line of thought is gaining attention: the accelerated
expansion is not a result of yet another ingredient in our already
gunky cosmic gas tank, but rather is a signal of our \modified{lack of
understanding of gravitational physics on large
scales}~\cite{Deffayet,Deffayet:2001pu,Damour:2002wu,Freese:2002sq,Freese:2002gv,Dvali:2003rk,Capozziello1,Carroll:2003wy,Capozziello2,Nojiri:2003ft,Khoury:2003aq,Khoury:2003rn}.

Being able to observationally differentiate the two possibilities,
dark energy versus modified gravity, is an essential component in
developing the modified-gravity paradigm.  One can easily envision
some modified-gravity model leading to an expansion history that can
be identically reproduced by some dark-energy model.  Thus,
observations that depend only on anomalous expansion histories are
insufficient to tease out the acceleration's root cause.  However, as
we argue in~\cite{Lue:2003ky}, if one attempts to modify cosmology at
today's Hubble scale, $H_0$, through altering the equations governing
gravitational dynamics, then generically one expects that the
gravitational force law of an isolated mass source is altered even at
distance scales much smaller than $H_0^{-1}$.  This effect can then be
exploited to differentiate between a modified-gravity explanation of
today's cosmic acceleration and dark energy (where the gravitational
force law remains unaltered)~\cite{Multamaki:2003vs,Lue:2003ky}.

A leading contender in modified-gravity explanations of acceleration
is the braneworld theory of Dvali, Gabadadze, and Porrati (DGP).  In
this theory, gravity appears four-dimensional at short distances but
\modified{is altered at distances large compared to some freely
adustable crossover scale $r_0$} through the slow evaporation of the
graviton off our four-dimensional braneworld universe into an unseen,
yet large, fifth
dimension~\cite{Dvali:2000hr,Dvali:2001gm,Dvali:2001gx}.  DGP gravity
provides an alternative explanation for today's cosmic acceleration
\cite{Deffayet,Deffayet:2001pu}: just as gravity is conventional
four-dimensional gravity at short scales ($r \ll r_0$) and appears
five-dimensional at large distance scales ($r \gg r_0$), so too the
Hubble scale, $H(t)$, evolves by the conventional Friedmann equation
at high Hubble scales but saturates at a fixed value as $H(t)$
approaches $r_0^{-1}$.  Thus, if one were to set
\modified{$r_0 \simeq H_0^{-1}$}, where $H_0$ is today's Hubble scale,
then DGP gravity could account for today's cosmic acceleration in
terms of the existence of extra dimensions and a modification of the
laws of gravity.  The resulting cosmic expansion history is specific
and may be tested using a variety of cosmological
observations~\cite{Deffayet:2001pu,branescan,Deffayet:2002sp,Alcaniz:2002qh}.
However, can we distinguish between DGP gravity and a dark energy
model that mimics the same cosmic expansion history?

We would naively expect not to be able to probe the extra dimension 
at distances much smaller than the crossover scale $r_0 = 
H_0^{-1}$. However, in DGP, although gravity is four-dimensional at distances
shorter than $r_0$, it is not four-dimensional Einstein gravity -- 
it is augmented by the presence of an ultra-light gravitational scalar. 
One only recovers Einstein gravity in a subtle fashion~\cite{Deffayet:2001uk,Lue:2001gc,Gruzinov:2001hp,Porrati:2002cp}, and a marked departure 
from Einstein gravity persists down to distances much shorter than $r_0$.  
For example, for $r_0\approx H_0^{-1}$ and a central mass source of Schwarzschild radius $r_g$, 
significant and cosmologically-sensitive deviations from Einstein gravity occur 
at distances greater than~\cite{Gruzinov:2001hp,Porrati:2002cp,Lue:2002sw,Dvali:2002vf}
\be
     r_* = \left(r_g r_0^2 \right)^{1/3}  \approx
     \left({r_g\over H_0^2}\right)^{1/3}\ .
\ee
Thus a marked departure from conventional physics persists down to scales 
much smaller than the distance at which the extra dimension is naively hidden, 
or for our discussion here, the distance at which the Friedmann equation was modified 
to account for accelerated cosmic expansion.  This alteration of gravitational interactions 
provides a way of differentiating between DGP gravity and dark energy models, 
and is consist with the argument we put forth in~\cite{Lue:2003ky}.

Imminent solar system tests have been 
\modified{shown to be capable}
of probing 
the residual deviation from four-dimensional Einstein gravity at distances well 
below $r_*$~\cite{Lue:2002sw,Dvali:2002vf}.
Nevertheless, it would be ideal to test gravitational physics where dramatic differences 
from Einstein gravity are anticipated.  A detailed study of large scale structure in the Universe 
can provide such tests of gravitational physics at large distance scales.  
Unfortunately, prior analyses related to modified-gravity explanations of cosmic 
acceleration~\cite{Multamaki:2003vs,Lue:2003ky} are not applicable here.  
The modified force law is, in effect, sensitive to the background cosmological expansion, 
since this expansion is intimately tied to the extrinsic curvature of the brane~\cite{Deffayet,Lue:2002fe}, 
and this curvature controls the effective Newtonian potential.  A more careful analysis must be performed. 
In the next section we briefly review DGP gravity and identify the new force law necessary to calculate 
how large scale structure evolves in this cosmological model.  We then proceed and compare those results 
to the standard cosmology, as well as to a cosmology that exactly mimics the DGP expansion history 
using dark energy. Finally, we discuss the observational implications of our results on 
the growth of structure in DGP gravity and conclude with some remarks and a discussion of future work needed 
to improve upon our treatment.

\section{DGP Gravity}

We review the important points of the DGP braneworld model, including
cosmology.  We then modify the calculation performed
in~\cite{Lue:2002sw} to do determine the gravitational force law in an
evolving cosmological background, rather than in a static background
deSitter space.  This calculation is the new result in this paper that
allows one to answer questions of cosmological interest developed in
the next sections.

\subsection{The DGP Model}

Consider a braneworld theory of gravity 
\modified{(one in which ordinary particles and fields, other than the graviton, are confined to a 
three-dimensional hypersurface -- the brane -- embedded in a higher dimensional space -- the bulk)}
with an infinite-volume bulk 
and a metastable brane graviton~\cite{Dvali:2000hr}.  
We take a four-dimensional braneworld embedded in a five-dimensional Minkowski spacetime.  
The bulk is empty; all energy-momentum is isolated on the brane.  The action is
\be
S_{(5)} = -\frac{1}{16\pi}M^3 \int d^5x
\sqrt{-g}~R +\int d^4x \sqrt{-g^{(4)}}~{\cal L}_m + S_{GH}\ .
\label{action}
\ee
The quantity $M$ is the fundamental five-dimensional Planck scale.  The first term in Eq.~(\ref{action}) corresponds to the Einstein-Hilbert action in five dimensions for a five-dimensional metric $g_{AB}$ (bulk metric) with Ricci scalar $R$.  The term $S_{GH}$ is the Gibbons--Hawking action.  In addition, we consider an intrinsic curvature term which is generally induced by radiative corrections by the matter density on the brane~\cite{Dvali:2000hr}:
\be
-\frac{1}{16\pi}M^2_P \int d^4x \sqrt{-g^{(4)}}\ R^{(4)}\ .
\label{action2}
\ee
Here, $M_P$ is the observed four-dimensional Planck scale (see~\cite{Dvali:2000hr,Dvali:2001gm,Dvali:2001gx} for details). Similarly, Eq.~(\ref{action2}) is the Einstein-Hilbert action for the induced metric $g^{(4)}_{\mu\nu}$ on the brane, $R^{(4)}$ being its
scalar curvature.  The induced metric is\footnote{Throughout this paper, we use $A,B,\dots = \{0,1,2,3,5\}$ as 	bulk indices, $\mu,\nu,\dots = \{0,1,2,3\}$ as brane spacetime indices, and $i,j,\dots = \{1,2,3\}$ as brane spatial indices.} 
\be
g^{(4)}_{\mu\nu} = \partial_\mu X^A \partial_\nu X^B g_{AB}\ ,
\label{induced}
\ee
where $X^A(x^\mu)$ represents the coordinates of an event on the brane labeled by $x^\mu$.  The action given by Eqs.~(\ref{action}) and~(\ref{action2}) leads to the following equations of motion
\be
{1\over 2r_0}G_{AB} + \delta({\rm brane})G_{AB}^{(4)}
= {8\pi\over \mpsq}T_{AB}|_{\rm brane}\ ,
\label{Einstein}
\ee
where $G_{AB}$ is the bulk Einstein tensor, $G_{AB}^{(4)}$ is the
Einstein tensor of the induced brane metric, and $T_{AB}|_{\rm brane}$
is the matter energy-momentum tensor on the brane. We have defined a
crossover scale
\be
	r_0 = {\mpsq \over 2M^3}\ .
\label{r0}
\ee
This scale characterizes that distance over which metric fluctuations propagating on the brane dissipate into the bulk~\cite{Dvali:2000hr}.

\subsection{The Cosmological Background}

Let us review some important details of the cosmological background, 
for a general expansion of a spatially-flat brane into a Minkowski-flat bulk.  
We are primarily interested in the late evolution of the
Universe, in particular the matter-dominated era where the energy-momentum content 
of the universe is well-represented by a pressureless distribution of galaxies, spatially homogeneous on the
largest scales.  The spatially homogeneous cosmological background of such a Universe 
is driven by energy-momentum given by
\be
T^A_B|_{\rm brane}= ~\delta (z)\ {\rm diag}
\left(\rho(\tau),0,0,0,0 \right)\ ,
\label{cosmo-EM}
\ee
with spacetime geometry dictated by the line element~\cite{Deffayet}
\be
ds^2 = \left(1\mp{\ddot{\bar a}\over\dot{\bar a}}|z|\right)^2d\tau^2
- \bar{a}^2(\tau)\left(1\mp{\dot{\bar a}\over\bar a}|z|\right)^2
\left[\delta_{ij}d\lambda^id\lambda^j\right]
- dz^2\ .
\label{cosmo1}
\ee
Here, dot refers to differentiation with respect to the cosmological time $\tau$, the coordinate $z$ is the extra dimension, and the brane scale factor, $\bar{a}(\tau)$ satisfies a modified Friedmann equation
\be
     H^2 \pm {H\over r_0} = {8\pi\over 3\mpsq}\rho(\tau)\ ,
\label{Fried}
\ee
where $H(\tau) = \dot{\bar{a}}/\bar{a}$.
The two choices of sign represent two distinct cosmological phases. 
The phase of interest (the self-accelerating phase) corresponds to the lower sign, 
but we keep both for the sake of completeness.

This new Friedmann equation Eq.~(\ref{Fried}) already makes the theory distinct 
from standard $\Lambda$CDM cosmology, and observational signatures constraining DGP cosmology 
have been considered in e.g.~\cite{Deffayet:2001pu,Deffayet:2002sp,Alcaniz:2002qh}. 
Using constraints from Type 1A supernovae \cite{Deffayet:2002sp}, 
the best fit $r_0$ is $r_0 = 1.21_{-0.09}^{+0.09}H_0^{-1}$, where
$H_0$ is today's Hubble scale.  Taking $H_0 \approx 70~{\rm km~s^{-1}Mpc^{-1}}$, it implies 
$r_0 \approx 5~{\rm Gpc}$.

However, we wish to focus on those properties of DGP gravity 
that are affected by the modification of the force law, 
and cannot be mimicked by some dark energy component that produced the same expansion
history.  Following this program, we focus particularly at distance scales much smaller 
than the Hubble radius, $H^{-1}$.  As described in the introduction, 
the gravitational force law is significantly different from four-dimensional Einstein, 
even at these short distance scales.  We wish to determine the form of the corrections in 
the background of the expected matter-dominated cosmology, Eq.~(\ref{Fried}).

We are concerned with processes at distances, $r$, such that $rH \ll 1$.  
Under that circumstance it is useful to change coordinates to a frame 
that surrenders explicit brane spatial homogeneity but preserves isotropy
\bea
r(\tau,\lambda^i) &=& \bar{a}(\tau)\lambda     \\
t(\tau,\lambda^i) &=& \tau + {\lambda^2\over 2} H(\tau)\bar{a}^2(\tau)\ ,
\eea
for all $z$ and where $\lambda^2 = \delta_{ij}\lambda^i\lambda^j$.
The line element becomes
\be
ds^2 = \left[1 \mp 2(H+\dot{H}/H)|z| - (H^2+\dot{H})r^2\right]dt^2
-  \left[1 \mp 2H|z|\right]\left[(1 + H^2r^2)dr^2 + r^2d\Omega\right] - dz^2\ ,
\label{cosmo2}
\ee
where here dot repreresents differentiation with respect to the new time coordinate, $t$.  Moreover, $H = H(t)$ in this coordinate system.  
All terms of ${\cal O}(r^3H^3)$ or ${\cal O}(z^2H^2,zHrH)$ and higher have been neglected.

The coordinate system in Eq.~(\ref{cosmo2}) will be the most useful when considering 
the cosmological scenarios we are interested in.  One can see that the bulk is a Rindler space.  
This has a fairly natural
interpretation if one imagines the bulk picture~\cite{Deffayet,Lue:2002fe}.
One imagines riding a local patch of the brane, which appears as hyperspherical surface expanding into 
(or away from) a five-dimensional Minkowski bulk.  This surface either accelerates or decelerates 
in its motion with respect to the bulk, creating a Rindler-type potential.

\subsection{Nonrelativistic Matter Sources}

We are interested in finding the metric for compact, spherically symmetric overdensities 
in the background of a matter-dominated cosmology. Because we are only concerned with distance scales such that
$rH \ll 1$, then to leading-order in $r^2H^2$ and $zH$, the solutions to the field equations Eqs.~(\ref{Einstein})
are also solutions to the static equations, i.e. the metric is quasistatic, 
where the only time dependence comes from the slow evolution of the extrinsic curvature of the brane.
To be explicit, we are looking at the nonrelativistic limit, where the metric,
or the gravitational potentials, of a matter source depends only on the
instantaneous location of its elements, and not on the motion of those
elements.

Under this circumstance, 
one can choose a coordinate system in which the cosmological metric 
respects the spherical symmetry of the matter source.
Let the line element be
\be
ds^{2} = N^2(t,r,z) dt^{2}
         - A^2(t,r,z)dr^2 - B^2(t,r,z)[d\theta^2 + \sin^2\theta d\phi^2]
	 -dz^{2}\ .
\label{metric}
\ee
We are interested in small deviations of the metric from flatness so we define new functions $\{n(t,r,z),a(t,r,z),b(t,r,z)\}$ such that
\bea
     N(t,r,z) &=& 1+n(t,r,z)     \nonumber \\
     A(t,r,z) &=& 1+a(t,r,z)     \\
     B(t,r,z) &=& r[1+b(t,r,z)]\ .     \nonumber
\eea
The key is that because we are interested primarily in phenomena 
whose size is much smaller than the cosmic horizon, 
the effect of cosmology is almost exclusively to control the extrinsic curvature, 
of the brane. 
\modified{This can be  interpreted as a modulation of the brane's stiffness  or 
 the strength of the scalar gravitational mode.}

We take the energy-momentum tensor to now be
\be
T^A_B|_{\rm brane}= ~\delta (z)\ {\rm diag}
\left(\rho(t)+\delta\rho(t,r),0,0,0,0 \right)\ ,
\label{EM}
\ee
where the source mass is an overdensity of compact support (i.e. its extent is some radius, $R\ll H^{-1}$). Given a source mass whose overdensity with respect to the cosmological background goes as $\delta\rho(r,t)$, we may define an effective Schwarzschild radius
\be
      R_g(r,t) = {8\pi\over \mpsq}\int_0^r r^2\delta\rho(r,t) dr\ .
\label{rg}
\ee
We solve the perturbed Einstein equations in quasistatic approximation 
by generalizing the method used in~\cite{Lue:2002sw}, 
obtaining the metric of a spherical mass overdensity $\delta\rho(t,r)$ 
in the background of the cosmology described by Eq.~(\ref{cosmo2}) (rather
than deSitter space).  The metric on the brane, using the residual gauge fixing $b(t,r)|_{z=0} = 0$, is then given by
\bea
rn'(t,r)|_{z=0} &=& {R_g\over 2r}\left[1+\Delta(r)\right] - (H^2+\dot{H})r^2
\label{brane-n}\\
a(t,r)|_{z=0} &=& {R_g\over 2r}\left[1-\Delta(r)\right] + {1\over 2}H^2r^2
\label{brane-a}
\eea
where dot now denotes differentiation with respect to $t$ and prime denotes differentiation with respect to $r$.  Note that the background contribution is included in these metric components.  The
quantity $\Delta(r)$ is defined as
\be
     \Delta(r) = {3\beta r^3\over 4 r_0^2R_g}
     \left[\sqrt{1+{8r_0^2R_g\over 9\beta^2r^3}}-1\right]\ ,
\label{Delta}
\ee
and
\be
     \beta = {1\pm2r_0H+2r_0^2H^2\over 1\pm 2r_0H}\ .
\label{beta}
\ee
Though it is not of explicit interest here, the full $z$--dependence of
the metric may be deduced from Eqs.~(\ref{brane-n}) and (\ref{brane-a})
using equations laid out in the appendix of Ref.~\cite{Lue:2002sw}
with trivial alterations accounting for the differing cosmological
background.

The result Eqs.~(\ref{brane-n})--(\ref{beta}) is valid for $r\ll
H^{-1}$ and $r\ll r_0$, but only if the spatial support of
$\delta\rho(r,t)$ extends only to radii much less than $H^{-1}$, so
that there is a clear distinction between the matter making up the
overdensity and the cosmological background.  The result is virtually
identical to the strictly static deSitter background case except there
$\beta = 1\pm 2r_0H$.  One may also confirm that in the absence of
perturbations (i.e.,  $\delta\rho$ or $R_g = 0$), the background metric
Eq.~(\ref{cosmo2}) is a consistent, quasistatic solution.  This point
is analogous to the well-known idea that one may reproduce the
Friedmann equation in matter-dominated cosmologies with just the
Newtonian interaction bewteen matter particles.

From inspection of Eq.~(\ref{beta}), we see that, in addition to $r_0$, there exists a new transition scale
\be
	r_* = \left[{r_0^2R_g\over\beta^2}\right]^{1/3}\ ,
\ee
such that when $r\ll r_*$, the Einstein phase, the metric functions on the brane reduce to
\bea
	n &=& -{R_g\over 2r} \pm \sqrt{R_gr\over 2r_0^2}
\label{Einstein-n}  \\
	a &=& {R_g\over 2r} \mp \sqrt{R_gr\over 8r_0^2}\ .
\label{Einstein-a}
\eea
When $r\gg r_*$, the weak-brane phase, the metric functions on the brane become
\bea
	n &=& -{R_g\over 2r}\left[1 + {1\over 3\beta} \right]
		- {1\over 2}(H^2+\dot{H})r^2
\label{weak-n}  \\
	a &=& {R_g\over 2r}\left[1 - {1\over 3\beta}\right] + {1\over 2}H^2r^2\ .
\label{weak-a}
\eea
In this phase, the extra scalar mode, the would-be radion, 
alters the effective Newton's constants for the gravitational potentials represented by $n(t,r)-n(t,r)|_{\rm background}$, 
the Newtonian potential, and $a(t,r)-a(t,r)|_{\rm background}$, the gravitomagnetic potential.

One may simply check that the full $(t,r,z)$-dependent metric satisfies 
the complete modified Einstein equations Eqs.~(\ref{Einstein}) to the desired order.  
Note that to this order of precision, the velocity of
the matter distribution $\delta\rho(t,r)$ does not affect spacetime geometry (until order $v^2$ or $vrH$), 
thus corroborating the quasistatic approximation.

\subsection{Caveats}

The approximations $v \ll 1$ ($v$ represents peculiar matter velocities)
and $rH \ll 1$ play a role in several places and allow a series of crucial
simplifications that need to be spelled out.  These two approximations
are lumped together because the Hubble-flow velocity and peculiar
velocities play almost identical roles in the relevant field equations.
The following are the operational simplifications:
\begin{itemize}
\item  Nonrelativistic, quasi-static sources.  Source-velocity dependent
contributions to the gravitational field are subleading.  One may use
the static Einstein equations and still be assured that the metric on the
brane is accurate to ${\cal O}(v^2,r^2H^2)$.
\item Near-field regime.  Related to the above simplification, the
source evolves slowly enough that radiative effects are negligible
at these radii.  We may safely avoid scalar radiation on the brane and
gravity-wave evaporation into the bulk and other radiative bulk
effects \cite{Deffayet:2002fn,Lue:2002fe}.  These radiative scalar
modes may also have classical instabilities that might become relevant
outside this regime \cite{Luty:2003vm}.
\item  Resolving background matter from perturbation.  This
simplification is specific to DGP gravity.  The metric components
Eqs.~(\ref{brane-n}) and (\ref{brane-a}) depends on the background
Hubble expansion, and the calculation crucially depends on the
assumption that the overdensity does not alter the background
cosmology.  This can been seen to be self-consistent in the DGP
field equations only when the radius of support for the
overdensity is much smaller than the Hubble radius, $H^{-1}$.
\item  Geodesic motion is Newtonian,  i.e. the geodesic equation
reduces to Newton's second law where the potential is $g_{00}$
or $n(t,r)$.
\end{itemize}
If we stray too far from the assumptions $v\ll 1$ and $rH \ll 1$, 
then effects safely disregarded may start intruding into and complicating the
analysis, particularly when $H\sim H_0$, introducing additional
effects of equal significance to the ones included here.

\section{Growth of Density Perturbations}

\subsection{Spherical Perturbations}

Let us consider the evolution of a spherical top-hat perturbation $\delta(t,r)$ 
\modified{of top-hat radius $R_t$},
where $\rho(t,r)=\bar{\rho}(t)(1+\delta)$ is the full density distribution 
and $\bar{\rho}(t)$ is the background density.  At subhorizon scales ($Hr \ll 1$),
the contribution from the Newtonian potential, $n(t,r)$, dominates the
geodesic evolution of the overdensity.  From Eq.~(\ref{weak-n}) it follows
the equation of motion for the perturbation $\delta$ is,
\be
\ddot{\delta} - \frac{4}{3} \frac{\dot{\delta}^2}{1+\delta}+2H\dot{\delta} = 4\pi G \bar{\rho}\, \delta (1+\delta) \left[ 1 + \frac{2}{3\beta} \frac{1}{\epsilon} \left( \sqrt{1+\epsilon}-1\right)\right]\ ,
\label{sphc}
\ee
\modified{where the definition of $\epsilon\equiv8r_0^2R_g/9\beta^2R_t^3$}
follows from the identification of the expression in square brackets with $1+\Delta(r)$ (see Eq.~(\ref{Delta})),
and we have restricted ourselves to the self-accelerating branch
(i.e., the lower sign choice in all equations in the previous section).

For clarity, we may recast the time evolution of $\beta$ and $\epsilon$ in
terms of $\delta$ and the {\em time-dependent} value of $\om$.
Defining $\om(t) \equiv {8\pi \bar{\rho}(t)\over 3M_P^2 H^2(t)}$, and
using the Friedmann equation, Eq.~(\ref{Fried}), 
\be
\beta = -\frac{1+\om^2}{1-\om^2},\ \ \ \ \  \ \ \ \ \ 
\epsilon = \frac{8}{9} \frac{(1+\om)^2}{(1+\om^2)^2}\ \om\delta\ .
\ee
We stress that $\om$ is a time-dependent quantity -- it goes to
unity at high redshift $1 \ll z \ll z_{eq}$, where the evolution is Einstein-deSitter
(but in the matter dominated regime), and at present it reduces to the usual
value that we denote as $\omt=\om(z=0)$.  We see that $\beta$ is
negative, of order unity at present, and approaches minus infinity at high
redshift, whereas $\epsilon$ is proportional to $\om \delta$ with a
coefficient of order unity. Note that for large $\delta$, Eq.~(\ref{sphc})
reduces to the standard evolution of spherical perturbations in general
relativity. However, when $\delta$ is small, the correction term in the
square brackets can be noticeably different from unity.

\subsection{Linear Growth}

Let us focus first on linear perturbation growth at scales $r\ll H^{-1}$.  In this regime $\delta(r,t) \ll 1$, therefore one is always in the weak-brane regime,\footnote{This is always true for top-hat perturbations, but in practice the size $r$ is related to the amplitude $\delta$ through the perturbation spectrum. However, for 1$\sigma$ fluctuations of scale $r=10-100$ Mpc/$h$ with typical profiles given by the two-point correlation function, $r_*$ corresponds to $5-15$ Mpc/$h$. Therefore perturbations accessible to large-scale structure surveys are a natural probe of DGP gravity in the weak-brane regime.} $r\gg r_*$, and the only effect of DGP gravity is a modification of
Newton's constant.  Equation~(\ref{sphc}) reduces to
\be
	\ddot{\delta} + 2H\dot{\delta} =
	4\pi G\bar{\rho} \left(1+{1\over 3\beta}\right)\delta\ .
\label{growth}
\ee
Note that the effective Newton's constant, 
\be
	G_{\rm eff} =G \left(1 + {1\over 3\beta}\right)\ ,
\label{Geff}
\ee
is time-dependent.  Since $\beta$ is negative, as time goes on the effective gravitational constant decreases, and this extra repulsion (compared to general relativity) leads to suppressed growth. For example, if $\omt=0.3$, $G_{\rm eff}/G=0.72,0.86,0.92$ at $z=0,1,2$.

\begin{figure}[t]
\centerline{\epsfxsize=10cm\epsffile{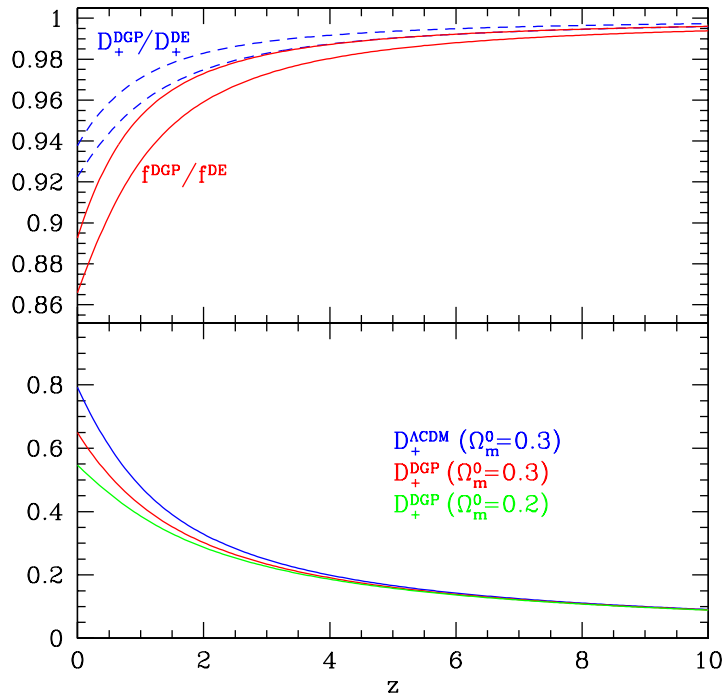}}
\caption{The top panel shows the ratio of the growth factors $D_+$ (dashed lines) in DGP gravity [Eq.~(\ref{growth})] and a model of dark energy (DE) with an equation of state such that it gives rise to the same expansion history (i.e. given by Eq.~(\ref{Fried}), but where the force law is still given by general relativity). The upper line corresponds to $\omt=0.3$, the lower one to $\omt=0.2$. The solid lines show the analogous result for velocity perturbations factors $f$. The bottom panel shows the growth factors as a function of redshift for models with {\em different} expansion histories, corresponding to (from top to bottom) $\Lambda$CDM ($\omt=0.3$), and DGP  gravity with $\omt=0.3,0.2$ respectively.}
\label{fig:growth}
\end{figure}

The growing-mode solution of Eq.~(\ref{growth}), $D_+$, is shown as a function of redshift $z$ in Fig.~\ref{fig:growth}. The top panel shows as dashed lines the ratio of $D_+$ in DGP gravity to that in a dark energy (DE) scenario with the same Friedmann equation but standard gravity, for two values of the present matter density $\omt=0.3$ (top) and $\omt=0.2$ (bottom). Notice how the change in the effective Newton constant leads to a suppression of $D_+$. Incidentally, this suppression is about two times larger than for models of modified gravity (with the same expansion history) that obey the Birkhoff's law~\cite{Multamaki:2003vs,Lue:2003ky}. The lower panel compares the growth factor $D_+$ to that in the standard cosmological constant scenario (with $\omt=0.3$ and $\Omega_\Lambda=0.7$), again for $\omt=0.3$ (top) and $\omt=0.2$ (bottom). We see here that the change in the expansion history (from a cosmological constant to DGP) leads to an additional suppression of the growth. In the language of dark energy, this is because the non-standard term in the  Friedmann equation Eq.~(\ref{Fried}) can be thought of as a contribution from a dark energy component with an effective equation of state given by 

\be
w_{\rm eff} = -\frac{1}{1+\om}\ ,
\label{weff}
\ee
therefore, for fixed $\omt$ such a term dominates the expansion of the universe earlier in DGP gravity  than in DE models with a cosmological constant, leading to an enhanced expansion rate $H$ and therefore and additional suppression over the one provided by the change in the force law. We will examine the observational consequences of this in Sec.~\ref{obs}. 

The growth of velocity perturbations is also a useful observable, 
and it follows directly from the continuity equation in the linear approximation. 
It is specified by $f \equiv d\ln D_+/d\ln a$ and it can be parametrized 
in terms of the time variable $\om$; 
in fact, one finds the following differential equation for $f(\om)$ directly from Eq.~(\ref{growth}), 
\be
-\frac{df}{d\om} + \frac{1}{3\om(1-\om)}\left[(2-\om) f + (1+\om)f^2\right] = 
	\frac{1}{3}\frac{(1+\om)(1+2\om^2)}{(1-\om)(1+\om^2)}
\label{fom}
\ee
whose numerical solution follows approximately $f(\om) \simeq \om^{2/3}$, 
which can be contrasted with the standard $f(\om) \simeq \om^{5/9}$ for flat models with a cosmological constant. 
The top panel of Fig.~\ref{fig:growth} shows the ratio of $f$ for DGP and DE models 
with the same expansion history, for $\omt=0.3,0.2$, 
showing that differences of {\em at least} $10\%$ are expected, 
whereas comparing DGP to cosmological constant models with the same $\omt$ larger differences are obtained, 
e.g. for $\omt=0.2$, $f^{\rm DGP}/f^{\rm \Lambda CDM}=0.83$. 
These deviations are well within the range that can be probed with current redshift surveys.

\subsection{Non-Linear Growth}
\label{nlgrowth}

\begin{figure}[t]
\centerline{\epsfxsize=10cm\epsffile{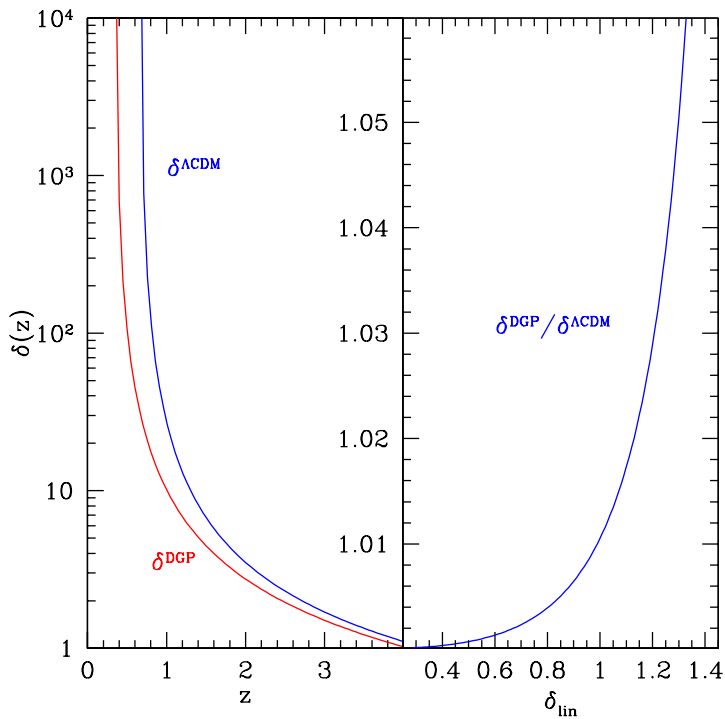}}
\caption{Numerical solution of the spherical collapse. 
The left panel shows the evolution for a spherical perturbation with $\delta_i=3\times 10^{-3}$ 
at $z_i=1000$ for $\omt=0.3$ in DGP gravity and in $\Lambda$CDM. 
The right panel shows the ratio of the solutions once they are both expressed 
as a function of their linear density contrasts.}
\label{fig:SphColl}
\end{figure}

The left panel in Fig.~\ref{fig:SphColl} shows the full solution of Eq.~(\ref{sphc}) 
with an initial condition of $\delta_i=3\times 10^{-3}$ at $z_i=1000$ for $\omt=0.3$, 
and the corresponding solution in the cosmological constant case. 
Whereas such a perturbation collapses in the $\Lambda$CDM case at $z=0.66$ 
when its linearly extrapolated density contrast is $\delta_c=1.689$, 
for the DGP case the collapse happens much later at $z=0.35$ when its $\delta_c=1.656$.  
In terms of the linearly extrapolated density contrasts things do not look very different, 
in fact, when the full solutions are expressed as a function of the linearly extrapolated density contrasts, 
$\delta_{\rm lin} = D_+ \delta_i/(D_+)_i$ they are very similar to within a few percent 
(right panel in Fig.~\ref{fig:SphColl}). This implies that all the higher-order moments of the density field 
will be very close to that in $\Lambda$CDM models. Indeed, such moments are determined 
by the vertices $\nu_n$ defined from ($\nu_1\equiv 1$)
\be
\delta(\delta_{\rm lin})= \sum_{n=1}^\infty \frac{\nu_n}{n!}\  \delta_{\rm lin}^n,
\label{vert}
\ee
e.g. the skewness is $S_3=3\, \nu_2$~\cite{Bernardeau92}, up to smoothing corrections that depend on the transformation from Lagrangian to Eulerian space. 
However since $\nu_n=d^n\delta/d\delta_{\rm lin}^n$ at $\delta_{\rm lin}=0$, 
the $\nu_n$'s in DGP gravity will all be very similar to those in $\Lambda$CDM 
(we have checked this explicitly for $S_3$, obtaining less than $1\%$ change). 
This can be useful in the sense that it allows the use of the non-linear growth to constrain 
the bias between galaxies and dark matter in the same way as it is done in standard case, thus inferring 
the linear growth factor from the normalization of the power spectrum in the linear regime. 
Although the result in the right panel in Fig.~\ref{fig:SphColl} may seem a coincidence at first sight, 
Eq.~(\ref{sphc}) says that the nontrivial correction from DGP gravity in square brackets is maximum 
when $\delta=0$ (which gives the renormalization of Newton's constant). 
As $\delta$ increases the correction disappears (since DGP becomes Einstein at high-densities), 
so most of the difference between the two evolutions happens in the linear regime, 
which is encoded in the linear growth factor.

\subsection{Late-time ISW Effect}

We now consider the late-time integrated Sachs-Wolfe (ISW) effect on the cosmic microwave background (CMB) 
for perturbations with scale $r\ll H^{-1}$. For this purpose, we need to identify
the gravitational potentials for linear overdensities as perturbations 
around a homogeneous cosmological background with the line element
\be
     ds^2 = \left[1+2\Phi(\tau,\lambda)\right]d\tau^2
     - \bar{a}^2(\tau)\left[1+2\Psi(\tau,\lambda)\right]
     \left[d\lambda^2+\lambda^2d\Omega\right]\ .
\label{potentials}
\ee
Here $\Phi(\tau,\lambda)$ 
and $\Psi(t,\lambda)$ are the relevant gravitational potentials and $\lambda$ is a comoving radial coordinate. 
In effect we want to determine $\Phi$ and $\Psi$ given $n$ and $a$. Unlike the case of Einstein's gravity, $\Phi \neq -\Psi$, due to additional contribution of the first term in Eq.~(\ref{Einstein}). 
One may perform a coordinate transformation to determine that relationship.  
We find that, assigning $r = \bar{a}(\tau)\lambda$, and
\bea
	\Phi &=& n - n|_{\rm background}	\\
	\Psi &=& -\int {dr\over r}\left[a(\tau,r) - a(\tau,r)|_{\rm background}\right]\ ,
\eea
keeping only the important terms when $rH \ll 1$.  But since we are concerned with linear density perturbations, we find from Eqs.~(\ref{weak-n}) and (\ref{weak-a}) that the quantity of interest for the ISW effect is the time derivative of 
\be
	\nabla^2(\Phi-\Psi) = {8\pi \over M_P^2}\bar{a}^2\rho\delta\ ,
	\label{poisson}
\ee
where $\nabla$ is the gradient in comoving spatial coordiantes. This result is identical to the four-dimensional Einstein result, the contributions from the brane effects exactly cancelling.  This result is not entirely surprising since the effect of the brane is the introduction of a new gravitational scalar that couples to the trace of the energy-momentum tensor.  However, the ISW effect has to do with the evolution of photons through a gravitational field (in the weak field limit), and photons will not couple to the gravitational scalar (its trace vanishes).  Thus, the late-time ISW effect for DGP gravity will be identical to that of a dark energy cosmology that mimics the DGP cosmic expansion history, Eq.~(\ref{Fried}), at least at scales small compared to the horizon. Our approximation does not allow us to address the ISW effect at the largest scales (relevant for the CMB at low multipoles), but it is applicable to the cross-correlation of the CMB with galaxy surveys.  At larger scales, one
expects to encounter difficulties associated with leakage of gravity off
the brane (for order-unity redshifts) and other bulk effects
\cite{Deffayet:2002fn,Lue:2002fe} that we were successfully able to
ignore at subhorizon scales.

Discussion of photon geodesics naturally leads one to ask how lensing
may be altered due to DGP contributions.  For weak lensing by large-scale structure, one is in the weak field limit and therefore Eq.~(\ref{poisson}) applies; that is,  the weak lensing pattern is
identical to that for Einstein gravity, apart from the difference in expansion histories and change in the force-law. In other words, reconstruction of the dark matter distribution in DGP from weak lensing only requires changing the growth rate and the geometrical distances.\footnote{In is a nontrivial result that
light deflection by a compact spherical source is identical to that in
four-dimensional Einstein gravity (even with potentials Eqs.~(\ref{brane-n})--(\ref{beta}) substantially differing from those of Einstein gravity) through the nonlinear transition between the Einstein phase and the weak-brane phase.  As such, there remains the possibility that for {\em aspherical} lenses that this
surprising null result does not persist through that transition and that DGP may manifest itself through some anomalous lensing feature.}

\subsection{Beyond Isolated Spherical Perturbations}

Since we have derived the growth of spherical isolated perturbations,
it is fair to ask how well do we expect our results to hold in the realistic
case of a superposition of perturbations of arbitrary shape. In the linear
regime, one expects to recover the same result as here, since the
linearized equations obey the superposition principle and one may
construct arbitrary perturbations from a linear superposition of
isolated spherical perturbations.  In the linear regime, DGP gravity
reduces to a Brans-Dicke theory with a slowly time-dependent
Newton's constant, Eq.~(\ref{Geff}).  The Newtonian potential is 
then just a solution to Poisson's equation for a given matter distribution
source.  It would be interesting, however, to corroborate this
prescription with a fully consistent linear solution for an arbitrary
perturbation spectrum along the lines of those presented
in~\cite{Deffayet:2002fn}, restricted to scales smaller than the
Hubble radius.

For the non-linear growth, the situation is more complicated since
the relation between the gravitational potential and the density
fluctuation $\delta$ is non-linear and the principle of superposition
no longer holds.  Here, however, one should keep in mind that these
non-linearities only develop late in the evolution after the universe
starts to accelerate, thus the corrections to superposition have a small
time to act. It is however difficult to say something more quantitative
at this point.

\section{Observational Consequences}
\label{obs}

What are the implications of these results for testing DGP gravity using large-scale structure?  
A clear signature of DGP gravity is the suppressed (compared to $\Lambda$CDM) growth of perturbations in the linear regime due to the different expansion history and the addition of a repulsive contribution to the force law.  However, in order to predict the present normalization of the power spectrum at large scales, we need to know the normalization of the power spectrum at early times from the CMB. A fit of the to pre-WMAP CMB data was performed in Ref.~\cite{Deffayet:2002sp} using the angular diameter distance for DGP gravity, finding a best fit (flat) model with $\omt\simeq 0.3$, with a very similar CMB power spectrum to the standard cosmological constant model (with $\omt\simeq 0.3$ and $\Omega_\Lambda^0=0.7$) and other parameters kept fixed at the same value. Here we use this fact, plus the normalization obtained from the best-fit cosmological constant  power-law model from WMAP~\cite{Spergel} which has basically the same (relevant for large-scale structure) parameters as in~\cite{Deffayet:2002sp}, except for the normalization of the primordial fluctuations which has increased compared to pre-WMAP data (see e.g. Fig.~11 in~\cite{Hinshaw}). The normalization for the cosmological constant scale-invariant model corresponds to present {\em rms} fluctuations in spheres of 8 Mpc/$h$,  $\sigma_8=0.9\pm 0.1$ (see Table 2 in~\cite{Spergel}). We assume a flat model, since it was shown in~\cite{Deffayet:2002sp} to be consistent as well with the DGP angular diameter distance. We ignore the fact that the ISW effect at low multipoles (where the CMB power spectrum has large error bars) can be different in DGP gravity, this has a small effect on the overall normalization of the primordial fluctuations that are determined by the overall power spectrum. 

\begin{figure}[t!]
\centerline{\epsfxsize=10cm\epsffile{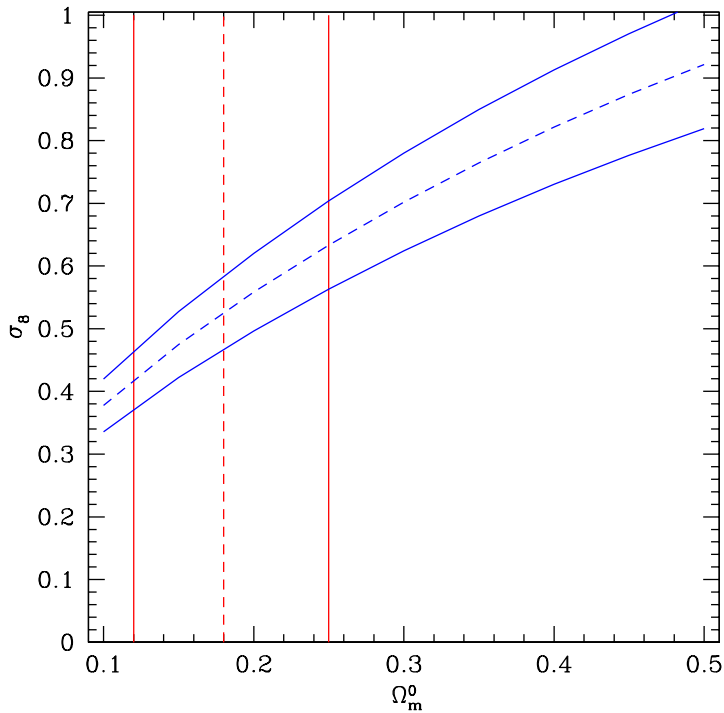}}
\caption{The linear power spectrum normalization, $\sigma_8$, for DGP gravity as a function of $\omt$. The vertical lines denote the best fit value and $68\%$ confidence level error bars from fitting to type-IA supernovae data from~\protect\cite{Deffayet:2002sp}, $\omt=0.18^{+0.07}_{-0.06}$. The other lines correspond to $\sigma_8$ as a function of $\omt$ obtained by evolving the primordial spectrum as determined by WMAP by the DGP growth factor. See text for details.}
\label{fig:sigma8}
\end{figure}

Figure~\ref{fig:sigma8} shows the present value of $\sigma_8$ as a function of $\omt$ for DGP gravity, where we assume that the best-fit normalization of the {\em primordial} fluctuations stays constant as we change $\omt$, and recompute the transfer function and growth factor as we move away from $\omt=0.3$. Since most of the contribution to $\sigma_8$ comes from scales $r<100 h$/Mpc, we can calculate the transfer function using Einstein gravity, since these modes entered the Hubble radius at redshifts high enough that they evolve in the standard fashion. The value of $\sigma_8$ at $\omt=0.3$ is then given by $0.9$ times the ratio of the DGP to $\Lambda$CDM growth factors shown in the bottom panel of Fig.~\ref{fig:growth}. The error bars in $\sigma_8$ reflect the uncertainty in the normalization of primordial fluctuations, and we keep them a constant fraction as we vary $\omt$ away from $0.3$. We see in Fig.~\ref{fig:sigma8} that for the lower values of $\omt$ preferred by fitting the acceleration of the universe, the additional suppression of growth plus the change in the shape of the density power spectrum drive $\sigma_8$ to a rather small value. This could in part be ameliorated by increasing the Hubble constant, but not to the extent needed to keep $\sigma_8$ at reasonable values. The vertical lines show the best-fit and 1$\sigma$ error bars from fitting DGP gravity to the supernova data from~\cite{Perlmutter:1998np} in~\cite{Deffayet:2002sp}. This shows that fitting the acceleration of the universe requires approximately $\sigma_8\leq0.7$ to 1$\sigma$ and $\sigma_8\leq0.8$ to 2$\sigma$. 

In order to compare this prediction of $\sigma_8$ to observations one must be careful since most determinations of $\sigma_8$ have built in the assumption of Einstein gravity or $\Lambda$CDM models. We use galaxy clustering, which in view of the results in Sect.~\ref{nlgrowth} for higher-order moments, should provide a test of galaxy biasing independent of gravity being DGP or Einstein. 
Recent determinations of $\sigma_8$ from galaxy clustering in the SDSS survey~\cite{Tegmark03a} give $\sigma_8^*=0.89\pm 0.02$ for $L^*$ galaxies at an effective redshift of the survey $z_s=0.1$. We can convert this value to $\sigma_8$ for dark matter at $z=0$ as follows. We evolve to $z=0$ using a conservative growth factor, that of DGP for $\omt=0.2$. In order to convert from $L^*$ galaxies to dark matter, we use the results of the bispectrum analysis of the 2dF survey~\cite{Verde02} where $b=1.04\pm0.11$ for luminosity $L\simeq 1.9L^*$. We then scale to $L^*$ galaxies using the empirical relative bias relation obtained in~\cite{Norberg01} that $b/b^*=0.85+0.15(L/L^*)$, which is in very good agreement with SDSS (see Fig.~30 in~\cite{Tegmark03a}). This implies $\sigma_8=1.00\pm0.11$. Even if we allow for another $10\%$ systematic uncertainty in this procedure, the preferred value of $\omt$ in DGP gravity  that fits the supernovae data is about 2$\sigma$ away from that required by the growth of structure at $z=0$. 

An independent way of testing DGP gravity with large-scale structure is to constrain the growth of velocity fluctuations through $f$. This affects the redshift distortions of the power spectrum and can be extracted from measurements, though at present the errors are somewhat large (see e.g.~\cite{Tegmark03a}) for an accurate test, but this should improve soon. The interesting feature of this test is that it is independent of the normalization of the primordial fluctuations, unlike the $\sigma_8$ normalization discussed above; thus it will be important to check that the same conclusions follow in this case.

\section{Concluding Remarks}

In this paper we identified how one may test the modifications of the gravitational force law expected in Dvali--Gabadadze--Porrati (DGP) gravity at scales of cosmological interest.  While cosmology is altered when the Hubble scale becomes comparable to today's Hubble scale, $H_0^{-1}$, the force law is correspondingly altered at much shorter scales and affects, for example, the growth of density perturbations at redshifts of order unity.

Although the results obtained in this paper are qualitatively comparable to those found in~\cite{Lue:2003ky}, they differ in key ways since DGP gravity does not obey Birkhoff's law.  The form of the gravitational force law between localized mass sources is sensitive to the background cosmological expansion.  So, while the results show deviations from Newtonian gravity of order unity at distance
scales greater than the scale $r_* \sim (r_g/H_0^2)^{1/3}$, where $r_g$ is the Schwarzschild radius of the mass source, the quantitative details differ. In particular, the suppression of the growth of structure is a factor about two larger than in Birkoff-law models with the same expansion history. 
Moreover, DGP gravity deviates significantly from four-dimensional Einstein gravity through the emergence of an ultra-light graviscalar mode.  Because such a mode does not couple to photons, the
new effects do not manifest themselves in the late-time integrated Sachs--Wolfe (ISW) effect, at least at the subhorizon scales we consider.  While the gravitational potentials are indeed altered by order-unity factors late in the cosmic expansion history, they precisely cancel, so that the late-time ISW effect in DGP gravity is identical to that for a dark energy theory that mimics the DGP expansion history. The same situation applies to weak gravitational lensing.

We have done a first assessment of the observational viability of DGP gravity to simultaneously explain the acceleration of the universe and the growth of structure.  In order to improve the comparison against observations a number of issues remain unsolved. First, one would like to check that the linear growth factor for sub-horizon scales derived under the spherical approximation holds for more general perturbations, as expected by the superposition principle in the linear regime. A more non-trivial check would be to generalize this to the non-linear case, or at least second-order in perturbation theory, to check the deviations from superposition assumed here in the nonlinear case. To do a full comparison of the CMB power spectrum against data it remains to solve for the ISW effect at scales comparable to the horizon. Our treatment found no difference from general relativity (except from the change in the expansion history), but this is only valid at subhorizon scales.

Nevertheless, the main problem for DGP gravity to simultaneously explain cosmic acceleration and the growth of structure is easy to understand: the expansion history is already significantly different from a cosmological constant, corresponding to an effective equation of state with $w_{\rm eff}=-(1+\om)^{-1}$. This larger value of $w$ suppresses the growth somewhat due to earlier epoch of the onset of acceleration. In addition, the new repulsive contribution to the force law suppreses the growth even more, driving $\sigma_8$ to a rather low value, in contrast with observations. If as error bars shrink the supernovae results continue to be consistent with $w_{\rm eff}=-1$, this will drive the DGP fit to a yet lower value of $\omt$ and thus a smaller value of $\sigma_8$. For these reasons we expect the tension between explaining acceleration and the growth of structure to be robust to a more complete treatment of the comparison of DGP gravity against observations.

\begin{acknowledgments}
The authors thank C.~Deffayet, G.~Dvali, G.~Gabadadze, A.~Gruzinov, M.~Takada and
M.~Zaldarriaga for helpful communications and insights.  A.~L.  and
G.~S. wish to thank the CERN Theory Division and the Center for
Cosmology and Particle Physics for their hospitality.  This work is
sponsored by DOE Grant DEFG0295ER40898, the CWRU Office of the
Provost, NASA grant NAG5-12100, NSF grant PHY-0101738, and CERN.
G.~S. thanks Maplesoft for the use of Maple V.
\end{acknowledgments}

\end{document}